\documentclass[sensors,article,accept,moreauthors,pdftex]{Definitions/mdpi}

\usepackage{tabularx}
\usepackage{multirow} 
\usepackage{graphicx} 
\usepackage{tikz}
\usepackage{pifont}
\newcommand{\xmark}{\ding{55}}
\def\checkmark{\tikz\fill[scale=0.4](0,.35) -- (.25,0) -- (1,.7) -- (.25,.15) -- cycle;}

\usepackage{xcolor} 
\definecolor{blue}{RGB}{51,153,255}

\setitemize{parsep=6pt,itemsep=0pt,leftmargin=*,labelsep=5.5mm}
\setenumerate{parsep=6pt,itemsep=0pt,leftmargin=*,labelsep=5.5mm}
\setlist[description]{itemsep=0mm}

\usepackage[labelformat=simple]{subfig}

\firstpage{1}
\pubvolume{20}
\issuenum{22}
\articlenumber{6587}
\pubyear{2020}
\copyrightyear{2020}
\history{Received: 26 October 2020; Accepted: 16 November 2020; Published: 18 November 2020}

\Title{A Privacy-Preserving Healthcare Framework Using Hyperledger Fabric} 

\Author{Charalampos Stamatellis $^{1}$\href{https://orcid.org/0000-0002-3074-7506}{\orcidicon}, Pavlos Papadopoulos $^{1,}$*\href{https://orcid.org/0000-0001-5927-6026}{\orcidicon}, Nikolaos Pitropakis $^{1,2}$\href{https://orcid.org/0000-0002-3392-9970}{\orcidicon}, Sokratis~Katsikas $^{3,}$*\href{https://orcid.org/0000-0003-2966-9683}{\orcidicon} and William J. Buchanan $^{1}$\href{https://orcid.org/0000-0003-0809-3523}{\orcidicon}}

\AuthorNames{Charalampos Stamatellis, Pavlos Papadopoulos, Nikolakos Pitropakis, Sokratis Katsikas and William J. Buchanan}

\address{%
$^{1}$ \quad Blockpass ID Lab, School of Computing Edinburgh Napier University, Edinburgh EH10 5DT, UK; 40333763@live.napier.ac.uk (C.S.); N.Pitropakis@napier.ac.uk (N.P.); B.Buchanan@napier.ac.uk (W.J.B.)\\
$^{2}$ \quad Eight Bells LTD, Nicosia 2002, Cyprus\\
$^{3}$ \quad Department of Information Security and Communication Technology, Norwegian University of Science and Technology, 2815 Gjøvik, Norway}

\corres{Correspondence: pavlos.papadopoulos@napier.ac.uk (P.P.); sokratis.katsikas@ntnu.no (S.K.)}

\abstract{{Electronic health record (EHR) management systems require the adoption of effective technologies when health information is being exchanged. Current management approaches often face risks that may expose medical record storage solutions to common security attack vectors. However, healthcare-oriented blockchain solutions can provide a decentralized, anonymous and secure EHR handling approach. This paper presents PREHEALTH, a privacy-preserving EHR management solution that uses distributed ledger technology and an Identity Mixer (Idemix). The~paper describes a proof-of-concept implementation that uses the Hyperledger Fabric's permissioned blockchain framework. The proposed solution is able to store patient records effectively whilst providing anonymity and unlinkability. Experimental performance evaluation results demonstrate the scheme's efficiency and feasibility for real-world scale deployment.}}


\begin{document}

\section{Introduction}
\label{intro}

Traditional healthcare practices have often been influenced by the usage of paper-based medical records and these have evolved into electronic patient records~\cite{abbas2014review}. Thus, electronic health records (EHRs) often contain highly sensitive medical information, which is shared among healthcare providers, pharmacies and patients~\cite{dubovitskaya2017secure}. Current EHR management approaches include distributed or cloud server data storage; this practice can lead to diverse functionalities and financial complications. Further, according to Coventry and Branley~\cite{coventry2018cybersecurity}, healthcare organizations are vulnerable to a number of cybersecurity threats, including malware and~ransomware.

{Successful compromise of EHR management and storage providers can be achieved by deploying ransomware, which encrypts data until a ransom is paid, or~by hacking the EHR repository~\cite{coventry2018cybersecurity}.} The~WannaCry ransomware cryptoworm, used in the WannaCry cyber attack that resulted in a loss of £92 million, infected vulnerable servers and computers of the National Health Service (NHS) and encrypted medical data at 80 out of 236 NHS foundations~\cite{smart2018lessons}. Further, the~Medjack (Medical Device Hijack) cyber attack used malware that infected multiple unprotected medical devices, compromised network defences, breached user anonymity and accessed medical records~\cite{coventry2018cybersecurity}. Healthcare information is often substantially more valuable than other data sources for trading on the black market: the~average cost of a hijacked medical record is approximately \$380; this is twice as large as the average cost across all industry-related data breaches~\cite{coventry2018cybersecurity}. {Incidents such as the above demonstrate that the healthcare sector can suffer substantial financial loss as a result of common attack vectors against traditional EHR databases~\cite{alvarez2017security}}.

With the increasing drive towards EHR, there is a certain urgency for a scalable, immutable, transparent and secure solution to be implemented in order to address the aforementioned challenges. A~state-of-the-art approach, invulnerable to common attack vectors, could handle medical data in a decentralized manner in order to avoid the possibility of a single point-of-failure or a single point-of-attack. A~key driver is to achieve general agreement amongst the co-operating healthcare providers in order for weak security and insider threats to be prevented or properly responded~to.

Current EHR storage and distribution approaches utilize certain solutions in the context of challenging the aforementioned security threats. In~particular, they use access control policies such as role-based access control (RBAC) and attribute-based access control (ABAC), which restrict system access to unauthorized users according to specific preassigned roles and attributes~\cite{yuksel2017research}. {Moreover, a~secure and anonymous three-factor authentication protocol could be deployed in the context of privacy-preserving healthcare-oriented wireless sensor networks (WSNs) \cite{renuka2019cryptanalysis}}. Additionally, encryption combined with pseudonymization techniques is deployed to conceal users' identities and preserve anonymity. However, access control policies may leak private information and encryption mechanisms could potentially affect system performance~\cite{dubovitskaya2015cloud}.

\subsection*{Contributions}
\begin{figure}[H]
\centering
\includegraphics[width=0.7\linewidth]{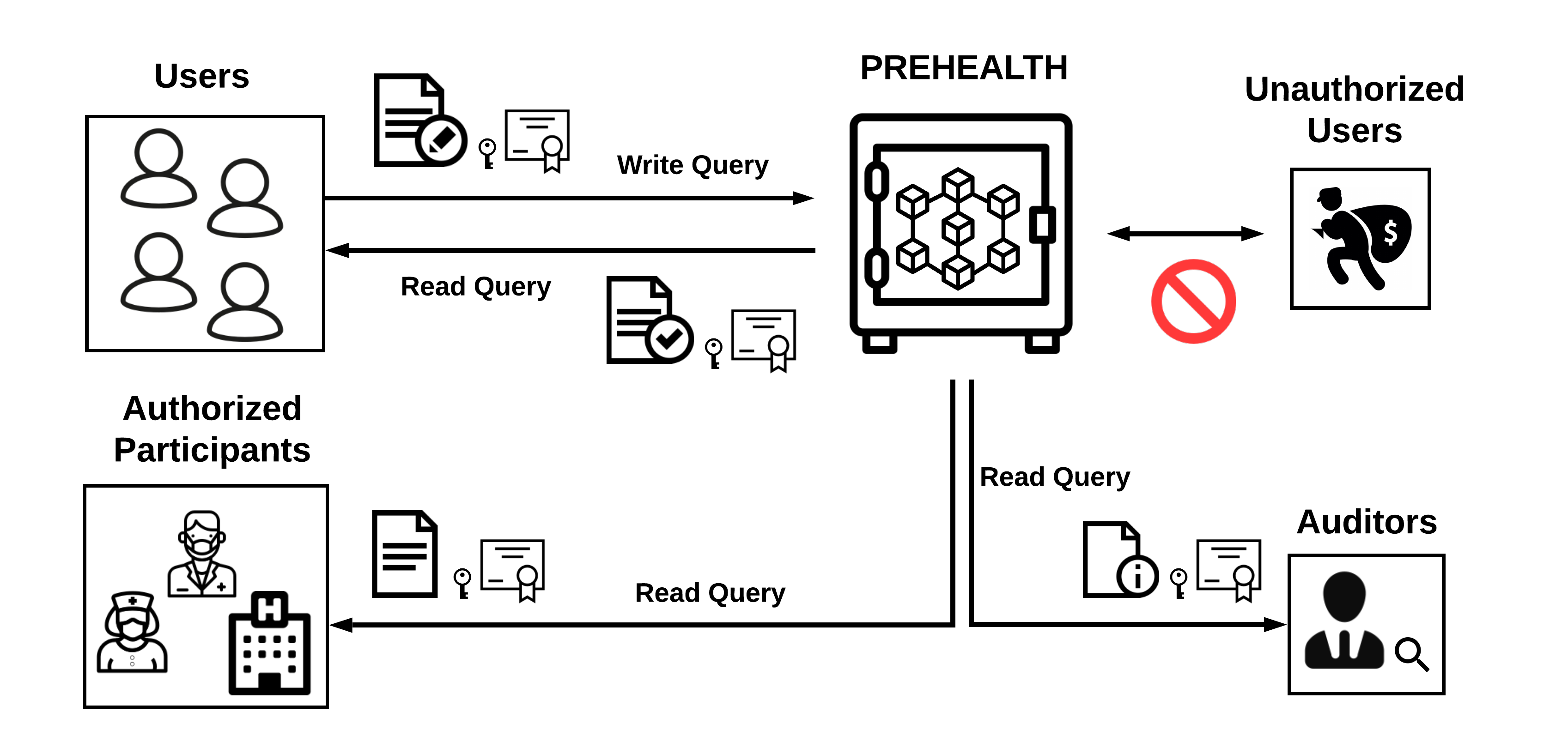}
\caption{{Privacy-Preserving Healthcare overview}.}
\label{fig:pphoverview}
\end{figure}

{In this paper we propose Privacy-Preserving Healthcare (PREHEALTH), a~privacy-preserving EHR management solution, illustrated in Figure~\ref{fig:pphoverview}. PREHEALTH can be used by patients or other healthcare actors that wish to store health records in an immutable distributed ledger that ensures data privacy. This immutable ledger protects data against misuses and abuses, whilst allowing the querying of non-personal data by authorized blockchain participants, for~example medical professionals. Additionally, PREHEALTH supports secure auditing, preservation of the privacy of accumulated personal data and isolates data related to auditing and compliance checks.}

The contributions of our work can be summarized as~follows:
\begin{itemize}
\item The development of PREHEALTH, a~privacy-preserving healthcare data solution. This uses distributed ledger technology and the Identity Mixer (Idemix) suite. Idemix is a zero-knowledge proof (ZKP) cryptographic protocol that provides privacy-preserving features, such as anonymity and unlinkability.
\item An evaluation of the robustness and security of PREHEALTH against popular attack vectors.
\item Empirical comparison of the performance of PREHEALTH against relevant blockchain solutions in the literature, and~against a traditional database that offers related in-context
column-level encryption.
\end{itemize}

The Idemix suite is still under development and currently has certain limitations that would need to be improved in future releases, as~the utilization of the aforementioned state-of-art architecture within a health record distribution approach is an essential contribution. With~Idemix, a~client acting as a blockchain participant may read, update or share private medical records while preserving their anonymity and~privacy.

The remainder of the paper is organized as follows. Section~\ref{Background} provides a background on the adopted distributed ledger technology that is essential to assure the self-sustainability of the paper. The~related literature is discussed in Section~\ref{healthindustry}. In~Section~\ref{expenv}, the~proof-of-concept implementation of PREHEALTH is presented, which also serves as our evaluation testbed. Section~\ref{evaluation} presents the process and the results of the evaluation of PREHEALTH, and~ Section~\ref{discussion} discusses our findings. Finally, Section~\ref{conclusion} summarizes our conclusions and suggests items for future~work.

\section{Background}
\label{Background}

{Blockchain is one of the most disruptive technologies of recent years~\cite{de2018pbft}. The~term is used to describe a decentralized peer-to-peer network, where an append-only and auditable ledger is updated by non-trusting entities~\cite{kuo2017blockchain}. The~number of both validators and end-users of the platform can be specified (permissioned); nevertheless, the~public and anonymous (permissionless) access option is also possible.}

{A permissionless blockchain network on the Ethereum blockchain framework has been proposed to implement a digital couponing system. Efficient non-repudiation and decentralization have been achieved, but~scalability and anonymity lag behind~\cite{podda2020overview}. Longo, Podda and Saia~\cite{longo2020analysis} examined Subchains, a~contemporary blockchain technique, developed on the Bitcoin public ledger, which~allows third-party services to store data in their database and synchronizes them with the Bitcoin blockchain network. However, both synchronization and security considerations affect the anonymity and scalability of the proposed scheme.}

\label{hlfoverview}

On the other hand, an~example of a permissioned ledger is Hyperledger Fabric, which is an open-source blockchain platform that endorses strong security and identity features~\cite{cachin2016architecture}. Hyperledger Fabric introduces a novel architecture~\cite{androulaki2018hyperledger,thakkar2018performance} and determines where transactions are processed by executing a programming code (known as chaincode) written in Go, Java or JavaScript programming languages, in~the following execute-order-validate~flow:
\begin{enumerate}
\item \textbf{Execution Phase}. A~client application sends a transaction proposal to endorsing peers as specified by the relative endorsement policy, in~order to invoke a chaincode function with regard to interacting with the blockchain ledger. As~soon as the endorsers have successfully executed the chaincode, the endorsement is sent back to the client. A transaction is then assembled and signed with credentials obtained from a Membership Service Provider (MSP), which is a party that handles all the identities of peers and validators within each organization.
\item \textbf{Ordering Phase}. A~client then sends constructed transactions to the ordering service. This is a collection of nodes known as orderers, which effectively combine multiple transactions into a single block and then broadcast the ordered transactions to all peers in the network.
\item \textbf{Validation Phase}. Lastly, each peer verifies received transactions with regard to the endorsement policy and then updates the local ledger state.
\end{enumerate}

Specifically, Hyperledger Fabric demonstrates unique security mechanisms such as private data collections, which allow certain authorized participants to access only specific data~\cite{papadopoulos2020privacy}. To~store data, peers use local state databases~\cite{papadopoulos2020privacy,androulaki2018hyperledger} such as CouchDB, MongoDB and LevelDB~\cite{thakkar2018performance}. Participants in Hyperledger Fabric can be in the form of Docker containers. These containers~\cite{boettiger2015introduction} are a unit of software that packages programming code and dependencies in order for an application to be executed efficiently and reliably in a computing environment. Furthermore, applications are implemented in an isolated and safe surrounding, as~Docker provides strong security capabilities. Several technologies can be implemented on top of Hyperledger Fabric in order to enhance its security and privacy. This~includes the Idemix cryptographic protocol suite~\cite{androulaki2019endorsement}, {which uses ZKP to provide anonymity and unlinkability. Idemix is associated with identity certificates that each participant needs to use for performing each and every action on the distributed ledger.}

\section{Related~Work}
\label{healthindustry}

Electronic health records (EHR) often contain highly sensitive healthcare data, which are periodically distributed among healthcare providers, pharmacies and patients for clinical diagnosis and treatment~\cite{dubovitskaya2017secure}. Furthermore, critical medical information must be regularly updated and shared where proper consent is provided by the patient. Along with this we need strong availability, fast~access and~the appropriate encryption of these records~\cite{dubovitskaya2017secure,abramson2020distributed}.

There are currently several approaches regarding EHR management and how blockchain technology can be utilized to improve it~\cite{holbl2018systematic,mayer2020electronic}. Yüksel~et~al.~\cite{yuksel2017research} discussed the distinction of EHR storage between distributed and cloud design, and~propose that a centralized model could be established by applying the relevant decentralized settings and techniques of the cloud architectures. Overall,~cloud architecture refers to the structured storage and allocation of massive volumes of medical records among remote third-party service providers. Within~the Cloud, healthcare organizations and individuals are able to access the data by utilizing relevant security measures based on identity management, access control, policy integration and compliance management, thus accomplishing common requirements such as scalability, availability and cost-effectiveness~\cite{abbas2014review}.

{Sharma, Chen and Sheth~\cite{sharma2018toward} examined kHealth, a~practical privacy-preserving cloud-based system that handles health data acquired from \textls[-15]{Internet of Things (IoT) devices. Their system aims to build personalized health predictive models by applying efficient but computationally expensive privacy guarantees, since it employs various homomorphic encryption and differential privacy techniques. Meanwhile, it should be noted that scalability optimizations undermine privacy-protection~mechanisms.}}

{Dubovitskaya~et~al.~\cite{dubovitskaya2015cloud} introduced a scalable and privacy-preserving e-Health cloud system where medical data is encrypted through public-key cryptography over both local and cloud-stored databases, and~efficiently distributed by using access control policies specified by the patient. A~limitation of this implementation is the possible misconduct of trusted cloud server providers; these could violate anonymity by deducing private information from the user's IP address. They could also associate a pseudonym to a patient by launching an inference attack.}

{Casino~et~al.~\cite{casino2019systematic} addressed how blockchain technology could enhance several applications within the healthcare industry, such as the management of EHR, drug counterfeiting and~user-oriented medical research.}

{Ming and Zhang~\cite{ming2018efficient} proposed an efficient privacy-preserving access control (PPAC) scheme for EHR cloud-based systems which utilizes the cuckoo filter, a~novel attribute-based signcryption (ABSC) mechanism, to~achieve anonymity and computational efficiency. They provided extensive privacy guarantees and comparative performance evaluation results. However, compliance with the GDPR has not been investigated.}

Roehrs~et~al.~\cite{roehrs2017omniphr} distributed personal health record (PHR) information into data blocks. From~a~logical point-of-view, the~data storage seems centralized, but in~fact it is decentralized among the participating devices. The~authors noted that their proposed protocol \textit{openPHR} is feasible, extensible, elastic and can be adopted in practice by many organizations. Their architecture is presented in detail, but~the practicality of their work is being questioned. Additionally, the~authors mentioned that security and privacy are still lacking in their approach. It should be noted that a PHR is controlled by the patient in contrast to an EHR, which is controlled by a healthcare institution. However, both EHRs and PHRs are electronically stored and distributed and thus may be evaluated in terms of performance and scalability metrics, privacy-preserving features and GDPR~compliance.

{Approaches that use a similar backbone technology to ours include~\cite{ichikawa2017tamper,liang2017integrating}.} Ichikawa~et~al.~\cite{ichikawa2017tamper} presented solutions that utilize the permissioned blockchain Hyperledger Fabric, in~order to store healthcare records collected from mobile devices. However,~the authors built their model on the older v0.5 version of Hyperledger Fabric, which does not have the private data collection feature that the v1.4 version (that PREHEALTH is built upon) has. Moreover, their model does not support the Idemix suite to create the necessary privacy guarantees. Overall, their system is able to store data in an immutable ledger but without any privacy protection for end-users. It~should be noted that to incorporate the private data collection feature, an~update of their system is not possible without a complete re-design of their architecture. Likewise, Liang~et~al.~\cite{liang2017integrating} utilized Hyperledger Fabric to simulate a real-world scenario with different participating entities. In~their system, the~represented entities include users, wearable devices, healthcare providers, insurance companies, blockchain networks and cloud databases. However, later versions of Hyperledger Fabric introduced new challenges that their work needs to address, in~case of an architecture re-design and revision into a newer version that incorporates the private data collection feature. The~authors conclude with helpful metrics of the feasibility of each query on their system, although~important technical details are~missing.

MeDShare~\cite{xia2017medshare} has many concepts similar to our work, however the backbone blockchain framework is not explicitly selected. Moreover, the~authors focus more on discussing the fundamental building blocks of the blockchain technology, such as data blocks and smart contracts, than~on a proposed solution.

The MedRec software~\cite{azaria2016medrec} is a permissionless blockchain implementation, built~on~the Ethereum blockchain network, which demonstrates auditable and decentralized EHR management. It introduces both questionable transaction scalability and linkability likelihood between a transaction and an individual, thus compromising data and user anonymity. MedRec has inspired other authors to build upon it and try to mitigate its issues~\cite{yang2017blockchain}.

The MediBchain system~\cite{al2017medibchain} establishes a permissioned peer-to-peer approach and is designed on the Ethereum network. It uses a cloud-based server infrastructure to achieve acceptable scalability. However, even if data held by the participating parties remain in encrypted form, linkability remains an issue. Another concern is the transaction cost, which is not measurable. Another solution built on the Ethereum network is one that focuses on the pharmaceutical supply-chain: \textit{Modum.io} \cite{bocek2017blockchains}. In~this work, Bocek~et~al.~\cite{bocek2017blockchains} developed an architecture where the IoT devices collect temperature data, and~mobile devices get this data and connect to an HTTP server. The~HTTP server acts as a blockchain node and stores the data in PostgreSQL databases, using smart contracts. Their real-world scenario looks promising and can be adapted in an EHR use case, although~it is not decentralized in each stage. Where the HTTP server is hijacked by attackers or malfunctions, the~collected data is vulnerable and exposed. Subsequently, its usage in an EHR scheme is not advisable, considering the sensitive nature of EHR~data.

A significant and prominently distributed database technology that elaborates the blockchain technology is Blockstack~\cite{ali2017blockstack}. {Blockstack operates by default using the Gaia distributed database that is able to store its data decentralized in the users' web browsers instead of on a centralized web server, thus enhancing privacy. The~Blockstack framework is currently built on the Bitcoin blockchain, but~can be moved onto another platform.}

{Last, a~decentralized privacy-preserving blockchain solution built on a permissionless blockchain network, which combines cloud storage and medical data acquired from IoT devices, is examined in~\cite{dwivedi2019decentralized} in the context of developing a novel hybrid EHR management method. The~key advantages of the proposed solution include the utilization of lightweight symmetric and asymmetric key encryption in order to achieve effective anonymity and client authorization; however, performance efficiency and GDPR compliance have not been examined.}

\subsection*{PREHEALTH's~Features}
\label{prehealthfeatures}
{PREHEALTH is different from previous approaches as it enables an effective and secure handling operation of highly sensitive health data over Hyperledger Fabric's permissioned blockchain network. Hyperledger Fabric is suitable for managing health records as it provides a confidential, scalable and highly flexible infrastructure solution~\cite{albeyatti2018white}, by~virtue of the layers of permission and access control that it offers. Another advantage of PREHEALTH is that its privacy-preserving features are compliant with the GDPR~\cite{voigt2017eu}, in~particular as it supports the \textit{right to be forgotten}, by~means of levels of \textit{data anonymization}, and~also by allowing stored records to have a defined \textit{lifetime}. According to the established policy, when the defined \textit{lifetime} expires, records can be purged from the state database, and~only their hashes will remain on the immutable ledger as proof that the records existed. Similarly, participants who decide to delete their data can send a deletion transaction request to the blockchain network~\cite{davari2019access}. Likewise, a~hash proving that the data existed in the ledger would remain; however, correlation with the identity of the data-owner is computationally infeasible within reasonable time~limits. 
}

Comparisons of PREHEALTH against those works in the literature whose \textls[-15]{presentations include either proof-of-concept implementations that can be utilized in real-world scenarios or detailed proposals are shown in Table~\ref{tab:comparison}.} In~this table, the~underlying technology of each proposal is presented, alongside Checkmarks or Xmarks that indicate the effective privacy-preserving features, the~ compliance with the GDPR and~whether performance and scalability evaluation results such as those presented in Section~\ref{prehealthfeatures} have been provided.

\begin{table}[H]
\centering
\caption{{Comparison of related proposals}.}
\label{tab:comparison}
\scalebox{.95}[0.95]{\begin{tabular}{ccccccc}
\toprule
\textbf{Method} & \multicolumn{1}{l}{\textbf{Technology}} & \multicolumn{1}{l}{\textbf{Access}} & \multicolumn{1}{l}{\textbf{Verifiability}} & \textbf{Privacy-Preserving} & \textbf{GDPR} & \textbf{Performance/Scalability} \\ \midrule
~\cite{ming2018efficient} & AC Scheme & Private & Private & \checkmark & \xmark & \checkmark \\ \midrule
~\cite{bocek2017blockchains} & Ethereum & Private & Public & \xmark & \xmark & \xmark \\ \midrule
~\cite{al2017medibchain} & Ethereum & Private & Public & \xmark & \xmark & \checkmark \\ \midrule
~\cite{azaria2016medrec} & Ethereum & Open & \begin{tabular}[c]{@{}c@{}}Public/\\ Private\end{tabular} & \checkmark & \xmark & \xmark \\ \midrule
~\cite{ali2017blockstack} & \begin{tabular}[c]{@{}c@{}}Bitcoin/\\ Agnostic\end{tabular} & Open & Public & \checkmark & \checkmark & \xmark \\ \midrule
~\cite{xia2017medshare} & Agnostic & Open & Private & \xmark & \xmark & \checkmark \\ \midrule
~\cite{roehrs2017omniphr} & Peer-to-peer & Private & Private & \xmark & \xmark & \checkmark \\ \midrule
~\cite{liang2017integrating} & HLF & Private & Private & \xmark & \xmark & \checkmark \\ \midrule
~\cite{ichikawa2017tamper} & HLF & Private & Private & \xmark & \checkmark & \xmark \\ \midrule
Our work & HLF & Private & Private & \checkmark & \checkmark & \checkmark \\ \bottomrule
\end{tabular}}
\end{table}

\section{Experimentation~Environment}
\label{expenv}

\begin{figure}[H]
\centering
\includegraphics[width=0.55\linewidth]{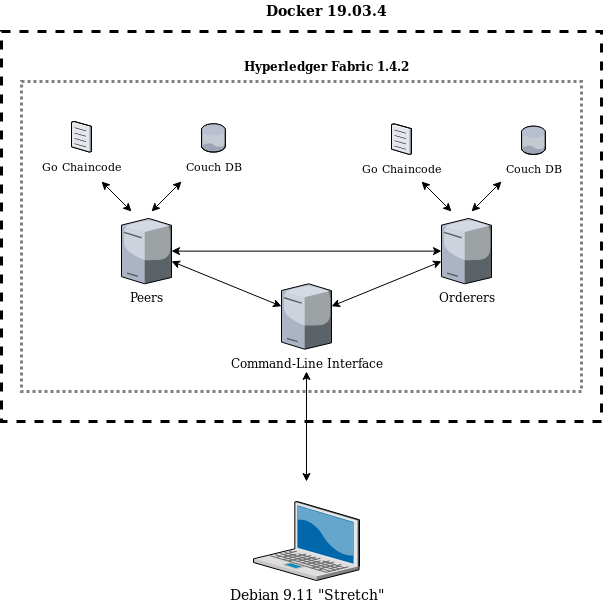}
\caption{Proof-of-concept implementation~overview.}
\label{fig:systemoverview}
\end{figure}

{The key focus of the work is to create a system that offers anonymity and security, and~acceptable scalability. To~this end, it was decided that multiple Docker containers, each incorporating a core element of Hyperledger Fabric's network, would be operated in a host environment, as~ illustrated in Figure~\ref{fig:systemoverview}. In~particular, a~Debian \textit{Stretch} host environment~\cite{krafft2010delphi} manages a dockerized command-line interface (CLI) in order to set up a Hyperledger Fabric network consisting of peers and orderers that have installed the chaincode and the distributed Couch DB ledgers~\cite{thakkar2018performance,androulaki2018hyperledger}.}

\begin{figure}[H]
\centering
\includegraphics[width=0.7\linewidth]{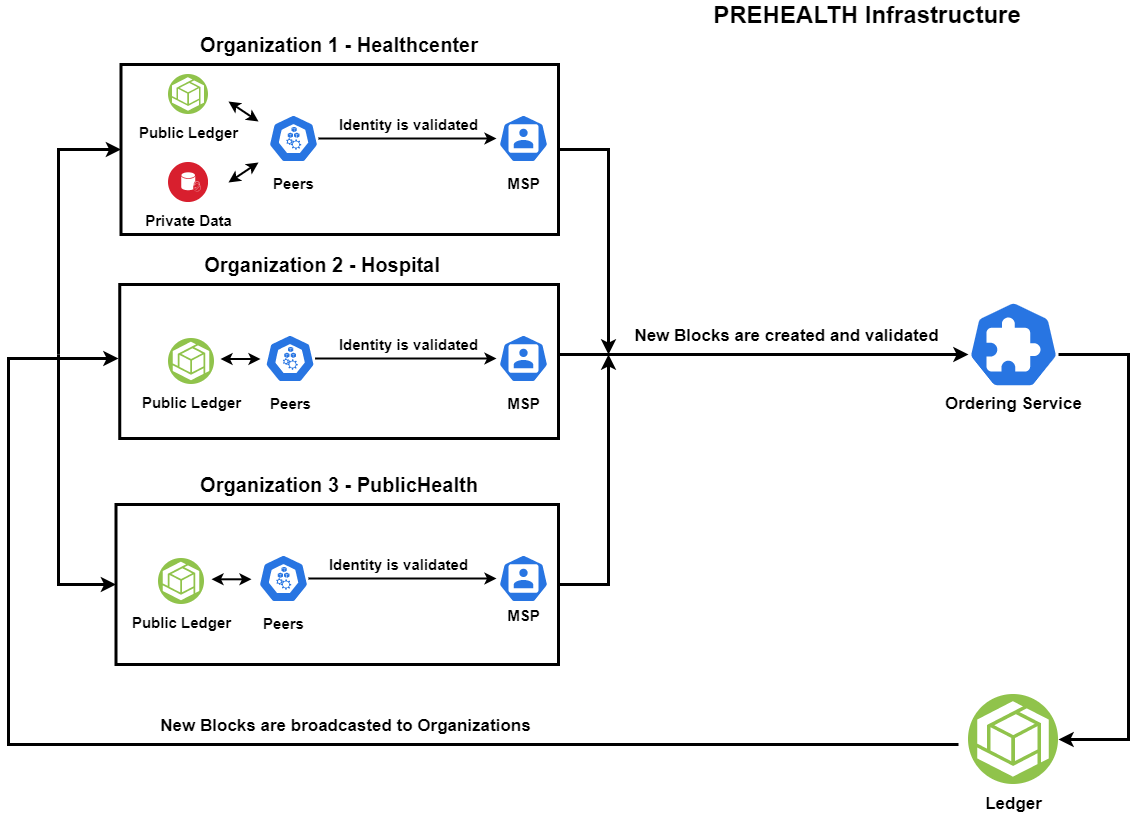}
\caption{{PREHEALTH overview}.}
\label{fig:infrastructureoverview}
\end{figure}

The overview for each distributed ledger action in Hyperledger Fabric is depicted in Figure~\ref{fig:infrastructureoverview}. Three different organizations are deployed, each consisting of three peers and a unique MSP. Each~peer maintains a copy of the public ledger, except~those from the peers of \textit{Organization 1} that manage a private data collection. This aims to demonstrate a scalable and privacy-preserving EHR management solution. As~described in Section~\ref{hlfoverview}, the~core functionality of the proposed testbed~is:

\begin{itemize}
\item Once a peer is validated by the MSP, it sends the proposed transaction to the ordering service.
\item The ordering service validates the transaction according to the associated chaincode and updates the public ledger.
\item Public ledger changes are broadcast to all the peers in order to verify, accept and update their local copy of the ledger.
\end{itemize}

The technical specifications of the proof-of-concept deployment are as follows: 2.4GHZ quad-core Intel Core i7 6th Generation CPU, with~8GB RAM and 256GB SSD. In~particular, each data record consists of six fields, of~which two represent sensitive patient data forming the private data collection and are encrypted. Specifically, the~fields are distinguished in the \textit{Name} of the patient, the~\textit{Address} of the patient, patient's \textit{Country} of residence, patient's \textit{Date of Birth}, a~\textit{Test} field that can include data related with each use case, and~finally a generated system-specific \textit{ID}. Moreover, the~private collection data consists of the first two fields, the~\textit{Name} and the \textit{Address} of the patient, and~is accessible only by \textit{Organization 1}, while \textit{Organizations 2 and 3} may access only the remaining non-sensitive data. Lastly, \textit{Read} and \textit{Write} query times of peers are measured as part of the evaluation of the proposed~solution.

\unskip

\section{Evaluation}
\label{evaluation}
\vspace{-6pt}

\subsection{Security~Evaluation}
\label{threatmodel}

It is common when it comes to technologies that involve identities and certificates, such as Hyperledger Fabric, that the~weakest link is the human; this is inherited from traditional IT systems. In~cases where the private certificates of a blockchain user are being disclosed or stolen, it is possible for a malicious user to perform arbitrary \textit{Read} and \textit{Write} queries to the blockchain ledger. According to the established policy, the~security of the system may be compromised. In~our case, this scenario is unrealistic, since no entity is completely authorized or has control over the ledger, not even the administrators. Furthermore, another risk for the blockchain systems may be the chaincode that is being executed in all the peers. Chaincode runs autonomously without external interventions, but~it may contain bugs. Extensive inspection and testing are crucial for the identification of bugs in the pre-production phase~\cite{androulaki2018hyperledger}. Last, as~the blockchain technology is still considered a new technology, while~offering new and robust security mechanisms that were impossible in the past, it is still possible that further bugs and attacks may be discovered in the future. One potential threat is quantum computing, which has been developed extensively over the last few years. All the approaches relying on hashing methods and the traditional encryption mechanisms may be at risk due to the aforementioned threat. Hence, a~proactive countermeasure is to use quantum-robust methodologies if they are practically feasible~\cite{Microsoftblockchain}.

\subsection{Anonymity~Evaluation}

\label{confparam}
The evaluation regarding the anonymity and unlinkability between state-of-art clients and transactions is focused on extracting relevant information during chaincode operation and blockchain user interaction. In~particular, disclosed identity attributes in both common user and Idemix-based client interactions are analyzed using the client identity library~\cite{CIDlibrary} embedded in the chaincode. It should be noted that the aforementioned library provides an exclusive method of acquiring private information about the validators as the defined communication channel utilizes the transport layer security (TLS) cryptographic protocol, thus efficiently encrypting relevant information, while~blockchain transactions are only signed by a cryptographic public-key, which does not reveal confidential data.

Furthermore, in~the work of~\cite{thakkar2018performance}, the~authors noted that Hyperledger Fabric provides a broad configuration space; therefore, particular parameters should be adjusted for anonymity and blockchain network performance measurements to be successfully determined and assessed. These components are divided into static and dynamic variables, and are~analyzed in the current section as~follows.
\vspace{6pt}

\textbf{Static variables:}
\begin{enumerate}
\item \textls[-15]{\textbf{Number of organizations, peers and orderers}---Considering complexity and technical limitations, three organizations were established, consisting of three peers each acting as validators and~an ordering service of three orderers.}
\item \textbf{Endorsement policy}---Each transaction proposal requires at least one peer of any corresponding organization to sign the transaction request, thus eliminating unnecessary communication between endorsing parties.
\end{enumerate}

\textbf{Dynamic Variables:}
\begin{enumerate}
\item \textbf{Membership service provider}---Idemix technology requires a distinct authentication provider in contrast to the prevailing X.509 public key certificate mechanism~\cite{chokhani1999rfc}.
\item \textbf{Registration scheme for users}---Unique CLI commands are required in the context of Idemix user registration. Specifically, a~suitable CLI docker container was developed, as~presented in Figure~\ref{fig:systemoverview}, in~which relevant Idemix parameters were handled as command-line arguments in order for a user to register and interact with the blockchain network.
\end{enumerate}

\subsection{Performance~Evaluation}
\begin{figure}[H]
\centering
\includegraphics[width=0.95\linewidth]{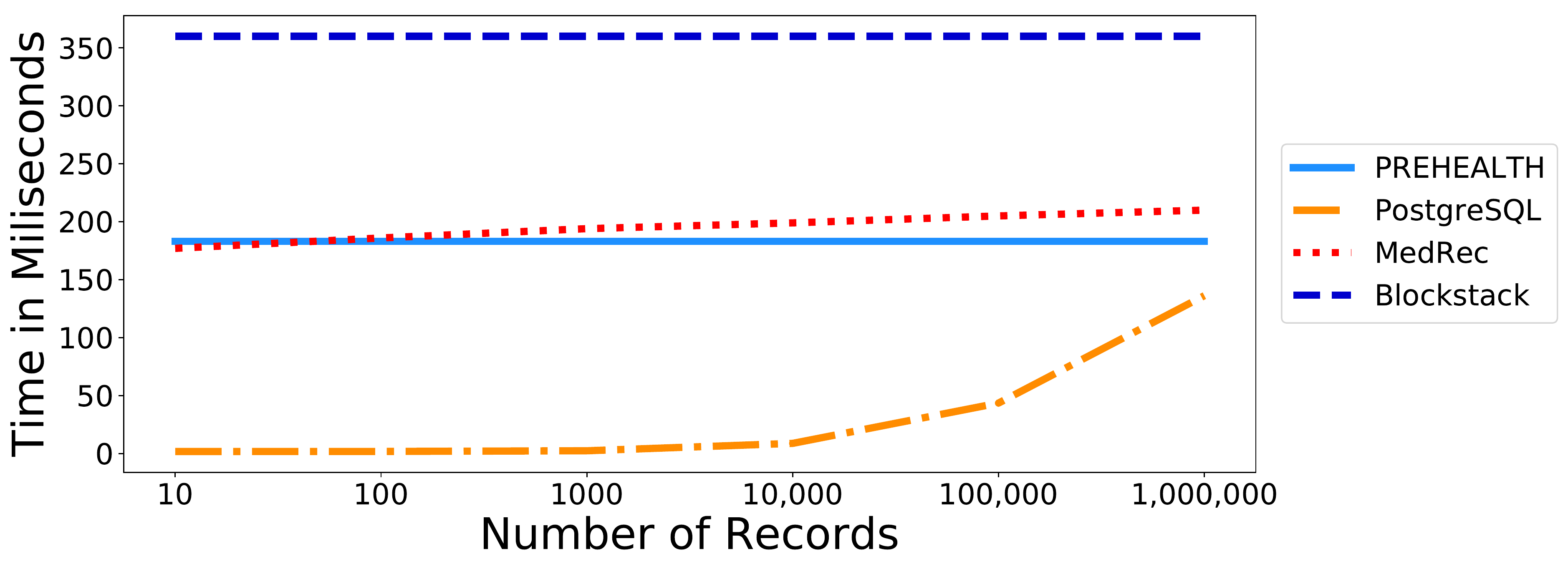}
\caption{{Read records transactions overhead}.}
\label{perf_fig}
\end{figure}

We calculated query times for PREHEALTH, as~well as for a PostgresSQL database simulating a common EHR management solution, and for the MedRec and Blockstack experimentation prototypes for various volumes of EHRs (10, 100, 1000, 10,000, 100,000 and 1,000,000). The~results are presented in Table~\ref{tab:metricscomparison} and in Figure~\ref{perf_fig}. PREHEALTH demonstrates efficient \textit{Read} and \textit{Write} query time measurements while both blockchain solutions, namely Medrec and Blockstack, but~also a traditional PostgreSQL database, impose higher overhead per query over time. Notably, PREHEALTH exhibits much better results when the number of records is large. As~can be seen from the conducted tests, the~transaction time for each \textit{Read} query using a PostgreSQL database is rising linearly and is expected to overcome PREHEALTH at approximately 1,200,000 records. It is clear, then, that in a real-world scenario where there are millions of stored records, PREHEALTH can handle each query effectively and~efficiently.

Furthermore, it should be noted that the MedRec prototype has been developed in an experimental environment by utilizing the Proof of Authority consensus algorithm, in~order for local nodes to be successfully deployed. This is in contrast to the operation of a time-consuming Proof of Work consensus model with regard to the commercial MedRec application~\cite{azaria2016medrec}, thus leading to increased query time~measurements.

\begin{table}[H]
\centering
\caption{Query time measurements in milliseconds (ms) per number of~entries.}
\label{tab:metricscomparison}
\scalebox{.95}[0.95]{\begin{tabular}{lccccccc}
\toprule
\multicolumn{2}{c}{\textbf{Number of Records:}} & {\textbf{10}} & \textbf{100} & \textbf{1000} & \textbf{10,000} & \textbf{100,000} & \textbf{1,000,000} \\ \midrule
\multicolumn{1}{l}{\multirow{2}{*}{PREHEALTH}} & \multicolumn{1}{c}{Read Data Time} & 183 ms & 183 ms & 183 ms & 183 ms & 183 ms & 183 ms \\ \cmidrule{2-8}
\multicolumn{1}{c}{} & Write Data Time & 58 ms & 58 ms & 58 ms & 58 ms & 58 ms & 58 ms \\ \midrule
\multirow{2}{*}{PostgresSQL Database} & Read Data Time & 1.73 ms & 1.79 ms & 2.38 ms & 8.76 ms & 43.52 ms & 136.19 ms \\ \cmidrule{2-8} & Write Data Time & 4.32 ms & 4.48 ms & 4.47 ms & 4.37 ms & 4.39 ms & 4.45 ms \\ \midrule
\multirow{2}{*}{MedRec---Azaria~et~al.~\cite{azaria2016medrec}} & Read Data Time & 177 ms & 186 ms & 194 ms & 199 ms & 205 ms & 210 ms \\ \cmidrule{2-8} & Write Data Time & 81.5 ms & 86.9 ms & 79.6 ms & 71.6 ms & 63.2 ms & 79.6 ms \\ \midrule
\multirow{2}{*}{Blockstack---Ali~et~al.~\cite{ali2017blockstack}} & Read Data Time & 360 ms & 360 ms & 360 ms & 360 ms & 360 ms & 360 ms \\ \cmidrule{2-8} & Write Data Time & 530 ms & 530 ms & 530 ms & 530 ms & 530 ms & 530 ms \\ \bottomrule
\end{tabular}}
\end{table}

Furthermore, benchmark tests were conducted for the \textit{CPU} and \textit{Memory} performance of our proposed PREHEALTH solution. The~CPU usage of PREHEALTH peers during the workflow for \textit{Read} and \textit{Write} queries over various volumes of EHRs (1000, 10,000, 100,000) is presented in Table~\ref{tab:average} and Figure~\ref{fig:cpuplots}. As~depicted in Figure~\ref{fig:infrastructureoverview}, PREHEALTH introduced three organizations, namely \textit{Healthcenter}, \textit{Hospital} and \textit{PublicHealth}. Each organization comprises three blockchain peers, respectively \textit{Peer 0}, \textit{Peer~1}~and \textit{Peer 2}. The~average CPU usage of the blockchain peers during \textit{Read} queries is reasonable and less than 30\%, whilst the CPU performance during \textit{Write} queries is even lower, with~a maximum of 15.4\%. The~CPU performance of the blockchain peers is regularly stable with swift fluctuations that are presumably caused by hardware limitations. The~average memory overhead of each blockchain peer during our tests over various volumes of EHRs was insignificant (less than 2\%); hence, it is not illustrated in a~plot.

\begin{table}[H]
\centering
\caption{Average CPU (\%) performance of each blockchain peer per number of electronic health~records.}
\label{tab:average}
\scalebox{.95}[0.95]{\begin{tabular}{cccccc}
\toprule
\multirow{2}{*}{\textbf{PREHEALTH Organizations}
} & \multicolumn{2}{c}{\multirow{2}{*}{\textbf{PREHEALTH Peers}}\vspace{-5pt}} & \multicolumn{3}{c}{\textbf{Number of Records}\vspace{2pt}} \\ \cmidrule{4-6}
& \multicolumn{2}{c}{}                   & \textbf{1000} & \textbf{10,000} & \textbf{100,000} \\ \midrule
\multirow{6}{*}{Healthcenter} & \multirow{2}{*}{Peer 0} & Read Queries & 7.6\%     & 28.7\%     & 29\%       \\
&                 & Write Queries & 6.7\%     & 10.3\%     & 15.4\%      \\ \cmidrule{2-6}
& \multirow{2}{*}{Peer 1} & Read Queries & 5.1\%     & 21.8\%     & 21.7\%      \\
&                 & Write Queries & 4.9\%     & 6.7\%      & 4.2\%      \\ \cmidrule{2-6}
& \multirow{2}{*}{Peer 2} & Read Queries & 4.9\%     & 23.3\%     & 22.2\%      \\
&                 & Write Queries & 5.4\%     & 6.4\%      & 4.3\%      \\ \midrule
\multirow{6}{*}{Hospital}   & \multirow{2}{*}{Peer 0} & Read Queries & 8.3\%     & 29.4\%     & 32.3\%      \\
&                 & Write Queries & 9.3\%     & 11.2\%     & 13.9\%      \\ \cmidrule{2-6}
& \multirow{2}{*}{Peer 1} & Read Queries & 5.1\%     & 22.7\%     & 23.2\%      \\
&                 & Write Queries & 5.4\%     & 6.4\%      & 4.3\%      \\ \cmidrule{2-6}
& \multirow{2}{*}{Peer 2} & Read Queries & 5.4\%     & 20.7\%     & 18.7\%      \\
&                 & Write Queries & 4.9\%     & 6.6\%      & 4.2\%      \\ \midrule
\multirow{6}{*}{PublicHealth} & \multirow{2}{*}{Peer 0} & Read Queries & 7.6\%     & 30.5\%     & 30.3\%      \\
&                 & Write Queries & 11.4\%     & 12.8\%     & 8.2\%      \\ \cmidrule{2-6}
& \multirow{2}{*}{Peer 1} & Read Queries & 4.8\%     & 22\%      & 20.1\%      \\
&                 & Write Queries & 5.3\%     & 6.8\%      & 4\%       \\ \cmidrule{2-6}
& \multirow{2}{*}{Peer 2} & Read Queries & 5.1\%     & 23.4\%     & 22\%       \\
&                 & Write Queries & 4.7\%     & 6.6\%      & 4\%       \\ \bottomrule
\end{tabular}}
\end{table}
\unskip

\begin{figure}[H]
\centering
\subfloat[]{{\includegraphics[width=7cm]{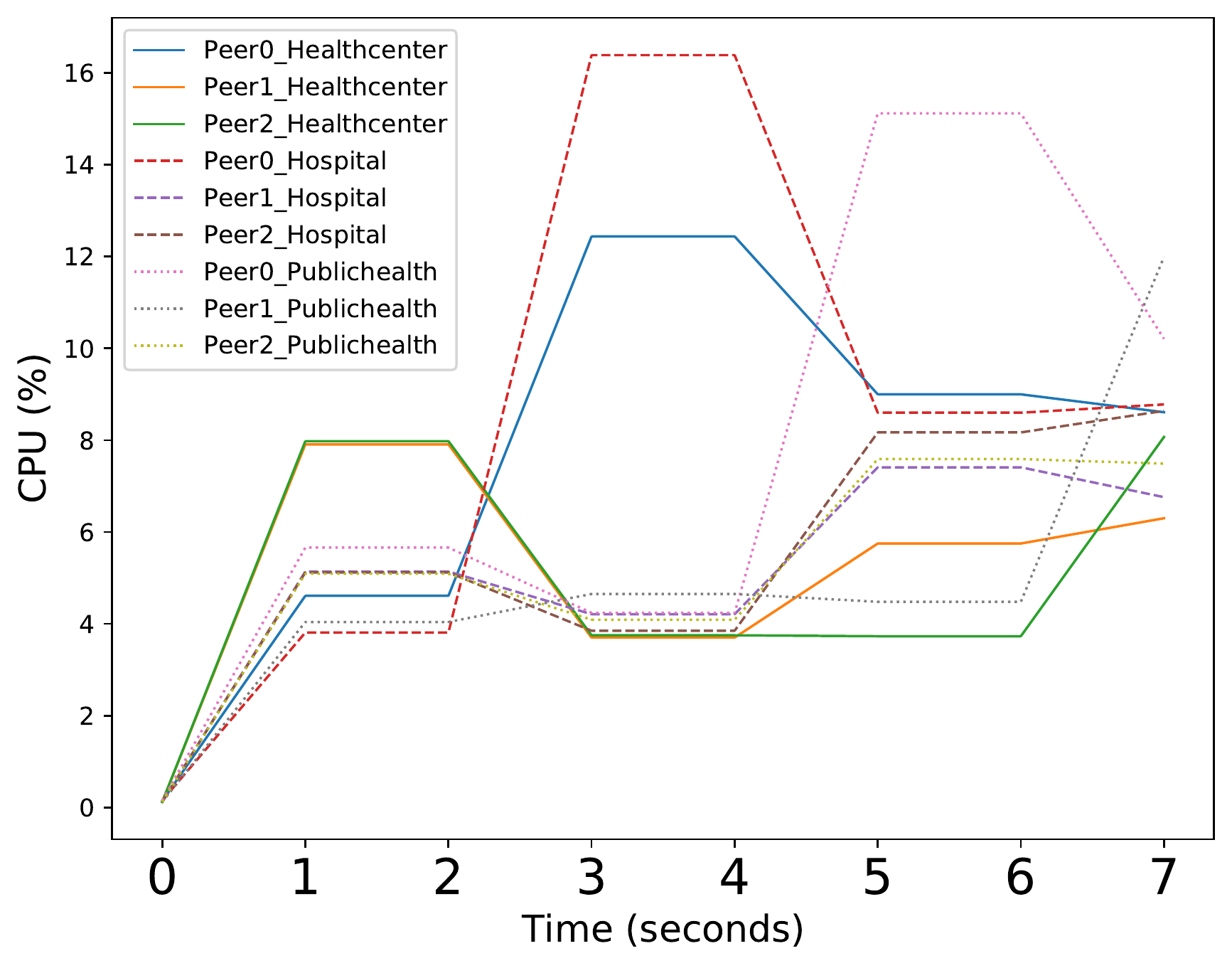} }}%
\qquad
\subfloat[]{{\includegraphics[width=7cm]{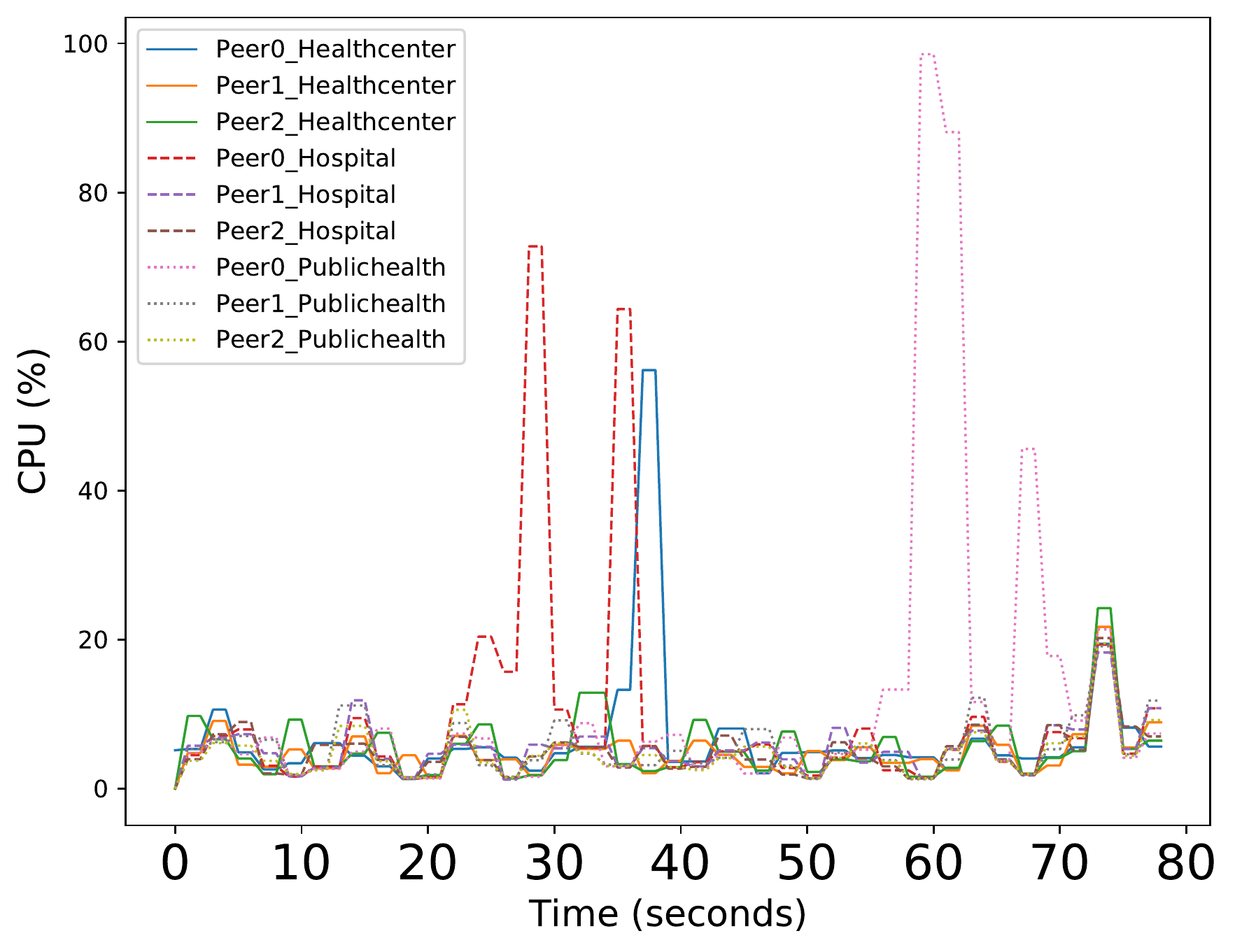} }}
\caption{\emph{Cont}.} 
\end{figure}
\begin{figure}[H]\ContinuedFloat
\centering
\subfloat[]{{\includegraphics[width=7cm]{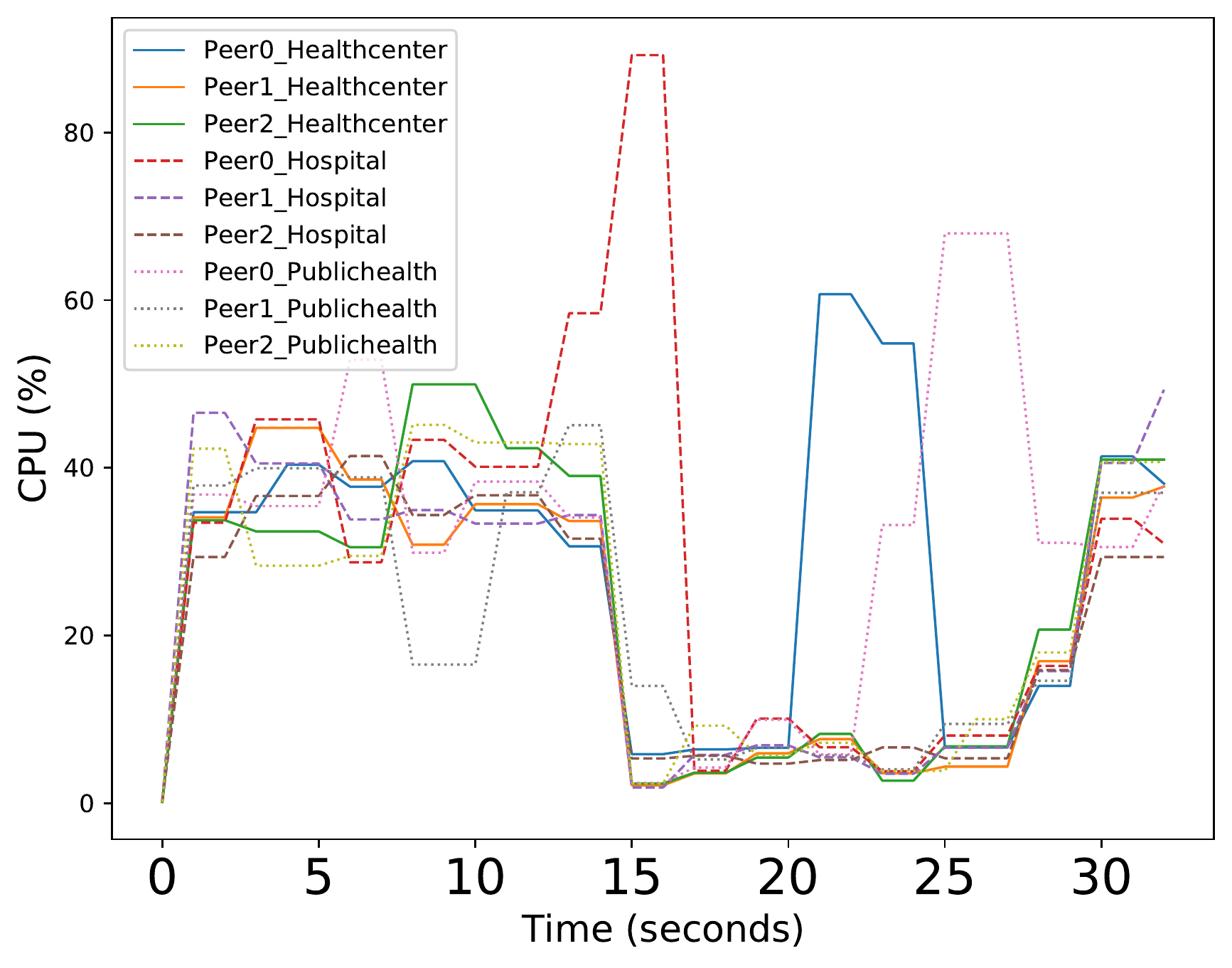} }}
\qquad
\subfloat[]{{\includegraphics[width=7cm]{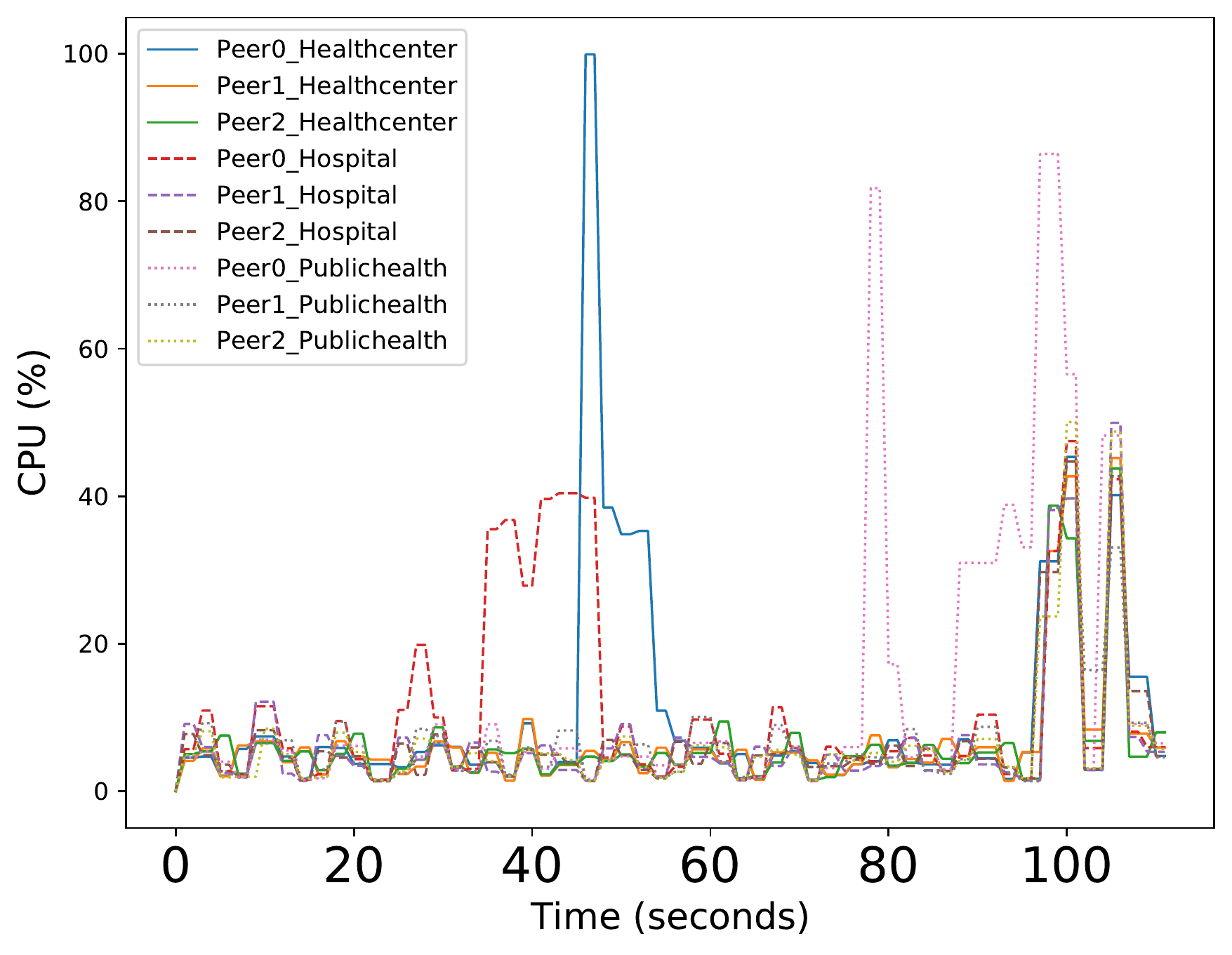} }}
\qquad
\subfloat[]{{\includegraphics[width=7cm]{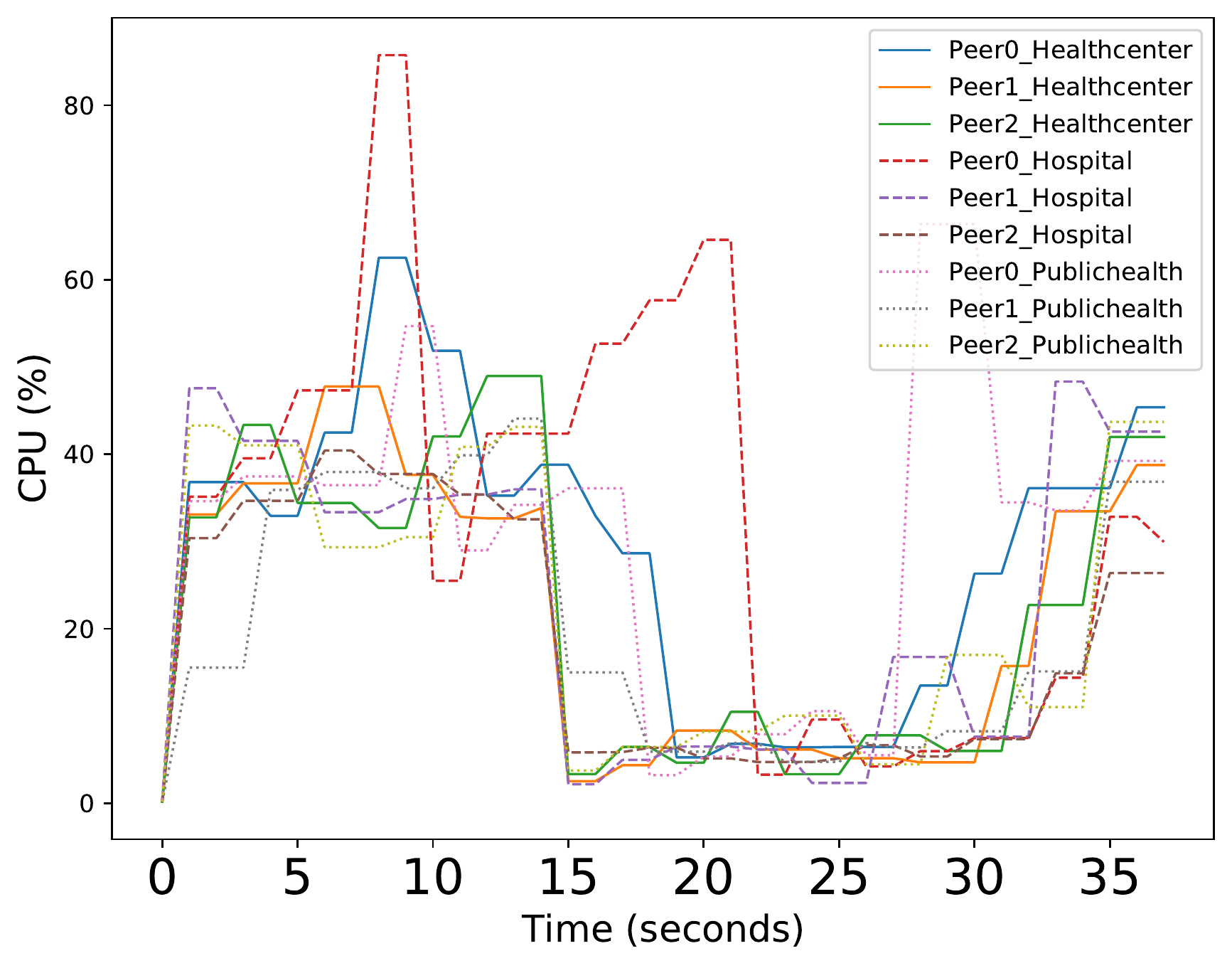} }}
\qquad
\subfloat[]{{\includegraphics[width=7cm]{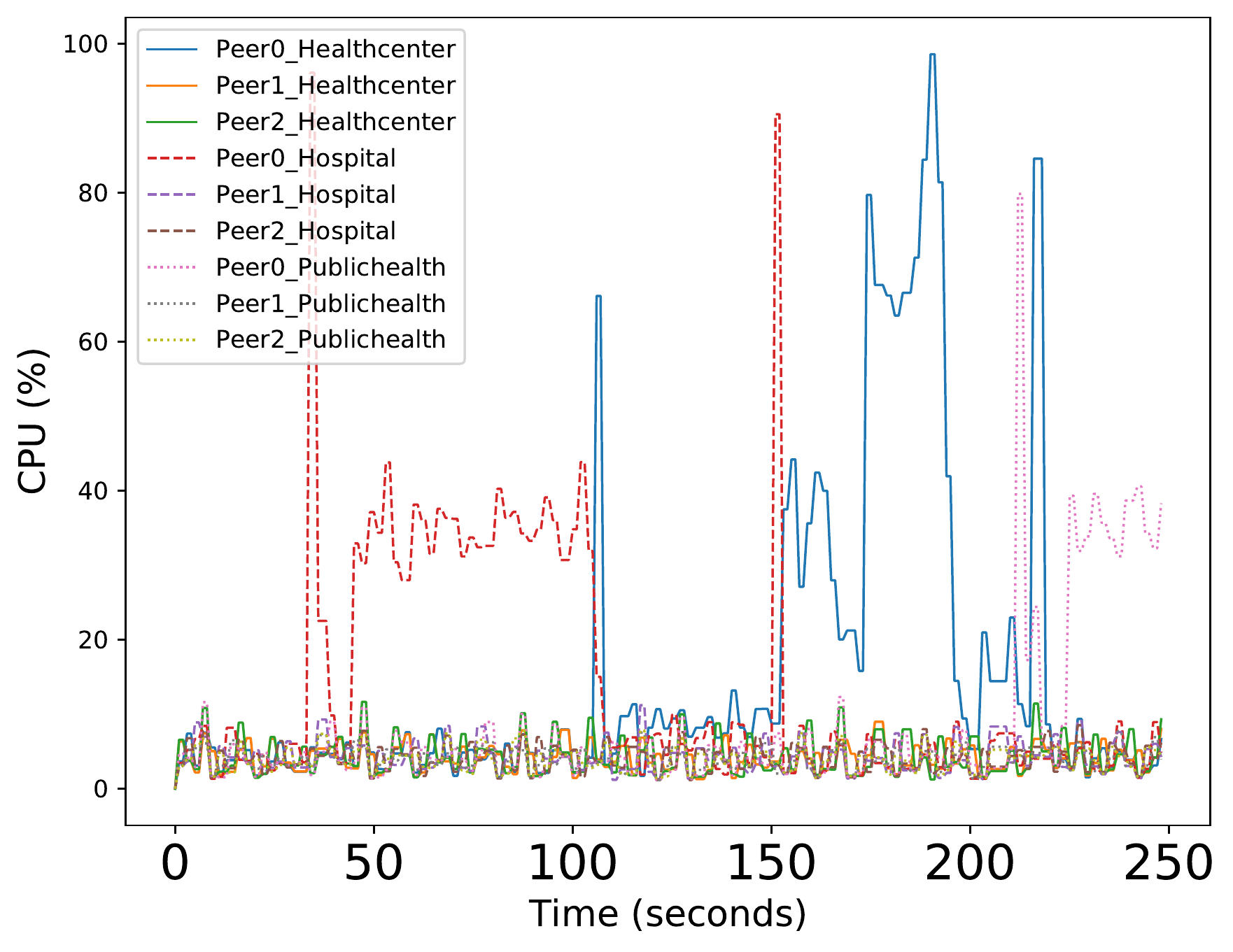}
}}
\caption{(\textbf{a}) {Read queries workflow on 1000 Records}. (\textbf{b}) {Write queries workflow on 1000 Records}. (\textbf{c})~{Read queries workflow on 10,000 Records}. (\textbf{d}) {Write queries workflow on 10,000 Records}. (\textbf{e}) {Read queries workflow on 100,000 Records}. (\textbf{f}) {Write queries workflow on 100,000 Records}. {CPU Usage (\%) of blockchain peers during workflow}.}
\label{fig:cpuplots}
\end{figure}

\section{Discussion}
\label{discussion}

An empirical comparison of PREHEALTH to the state-of-art procedures of the Hyperledger Fabric framework, which were utilized in order to achieve explicit anonymity and privacy of transactions, was conducted in the previous section. Additionally, experimental metrics in terms of blockchain performance, such as blockchain scalability, immutability, auditability and cost-effectiveness, were~assessed in association with selected environmental settings and parameters. Finally,~tests~were reviewed in terms of deploying an experimental proof-of-concept to a case-study~environment.

Calculation and examination of measurable parameters and performance metrics \textls[-15]{with regard to the experimental implementation were also considered. Apart from state-of-art Idemix technology utilization, certain blockchain features such as} scalability, auditability, immutability and cost-effectiveness were successfully validated. In~particular, our findings confirm that the proposed EHR management approach provides the capability for medical record transaction history to be both immutable and properly audited, even at a greater scale, without~significant performance~overhead.

Effective encryption mechanisms were applied to successfully encrypt information; as a result, a~malicious third party cannot access sensitive data. Furthermore, a~blockchain ledger is invulnerable to specific command injection threats such as SQL injection, which targets a specific format of an information database~\cite{halfond2006classification}. Moreover, the~MSP directory containing appropriate cryptographic material was mounted in each relative blockchain peer, thus rendering infeasible the exploitation of crafted cryptographic components by an attacker attempting to impersonate a blockchain peer and thus compromise the assembled network. Lastly, the~implementation was associated with a private isolated blockchain topology, invulnerable to external security threats, where identified blockchain peers and orderers would directly reveal their identity in the case of ill-behavior. In the case of a permissioned private blockchain consortium that would use PREHEALTH, all validator identities are disclosed. Therefore, upon~attempting to tamper with EHRs or blockchain transactions themselves, the~actor (i.e.,~validator) would reveal their~identity.

\section{Conclusions and Future~Work}
\label{conclusion}

The management of EHRs is and will continue to be crucial for years to come. EHRs consist of sensitive health data of subjects whose privacy must be protected. We proposed PREHEALTH, an~approach that can be used to store EHRs while ensuring the patients' privacy. We provided an empirical comparison of our proof-of-concept implementation against existing alternative proposals in the literature, and~found that PREHEALTH provides more privacy guarantees and at the same time imposes less query overhead when the volume of the stored data is high. Indeed, PREHEALTH achieves a read query time of 183 ms in any number of EHRs, when the time for a read query in the other compared works is higher or rising exponentially. Specifically, a~read query on the MedRec platform takes 210 ms in 1,000,000 EHRs, in~Blockstack it takes 360 ms and~a PostgreSQL database performs a read query in 136.19 ms. In~addition, we have proposed and used measurable parameters for evaluating the scheme's performance, scalability, auditability and immutability.

A possible extension of this work could be focused on operating an advanced version of Idemix credentials, in~a future release. Once credential revocation and incorporation of custom identity attributes are enabled, technical limitations of the Idemix technology would be resolved and a practical exploratory activity could be undertaken in order for a production privacy-preserving EHR handling solution to be developed. In~addition, another approach with regard to performance benchmarking could be a relative blockchain solution in a Kubernetes cluster, while a certain number of configurable parameters could be examined in association with their effects on blockchain scalability. Specifically, a~high-performance cloud environment could be utilized for the implementation of a complex blockchain network with the incorporation of numerous blockchain peers, orderers, endorsement policies and parallel blockchain interactions, thus efficiently evaluating metrics along with examining the scalability as a~whole.

\vspace{6pt}

\authorcontributions{All authors contributed to the conceptualization and methodology of the manuscript; C.S.~performed the data preparation; P.P. and N.P. contributed to writing; S.K. and W.J.B. reviewed and edited the manuscript. All authors have read and agreed to the published version of the~manuscript.}

\funding{The research leading to these results has been partially supported by the European Commission under the Horizon 2020 Program, through funding of the CUREX project (G.A. n 826404). }



\conflictsofinterest{The authors declare no conflict of~interest.}

\reftitle{References}
\externalbibliography{yes}


\end{document}